\documentclass[10pt]{book}

\usepackage[titles]{tocloft}

\usepackage{amsmath,amssymb}
\usepackage[Sonny]{fncychap}
\usepackage[pdftex]{geometry}

\usepackage{makeidx}

\usepackage{rotating}

\usepackage{graphicx}
\usepackage{pst-plot}
\usepackage{pst-node,pst-coil}
\usepackage{mathrsfs}

\usepackage{pst-3dplot}

\usepackage{epsfig}

\makeindex

\usepackage[hang,splitrule]{footmisc}



\usepackage{fancyhdr}
\begin{document}

\def\cs#1#2{#1_{\!{}_#2}}
\def\css#1#2#3{#1^{#2}_{\!{}_#3}}
\def\ocite#1{[\onlinecite{#1}]}
\def\ket#1{|#1\rangle}
\def\bra#1{\langle#1|}
\def\expac#1{\langle#1\rangle}
\def\dbl{\hbox{${1\hskip -2.4pt{\rm l}}$}}
\def\bfh#1{\bf{\hat#1}}

\newenvironment{rcase}
    {\left.\begin{aligned}}
    {\end{aligned}\right\rbrace}

\makeatletter
\renewcommand{\DOCH}{%
\vspace*{-80\p@}
\raggedleft
    \CNV\FmN{\@chapapp}\space \CNoV\thechapter
    \par\nobreak
    \vskip 40\p@}
\renewcommand{\DOTI}[1]{%
\CTV\raggedleft\mghrulefill{\RW}\par\nobreak
    \vskip 0\p@
    \CTV\FmTi{#1}\par\nobreak
    \vskip -10\p@
    \mghrulefill{\RW}\par\nobreak
    \vskip 22\p@}
\renewcommand\chapter{\par%
  \thispagestyle{plain}%
  \global\@topnum\z@
  \@afterindentfalse
  \secdef\@chapter\@schapter}
\renewcommand*\@makeschapterhead[1]{%
{\parindent \z@ \raggedright
\normalfont \huge\bfseries
#1\par\nobreak
\vskip 0\p@}}
\makeatother

\ChNameVar{\sl}

\ChNameAsIs

\ChTitleVar{\LARGE\bf\centering}

\ChNumVar{\fontsize{60}{62}\usefont{OT1}{pzc}{m}{n}\selectfont}

\parskip 0.17cm

{\renewcommand{\thefootnote}{\fnsymbol{footnote}}

\begin{titlepage}
\title{\Huge On the Origins of Quantum Correlations}
%\title{\Huge From Einstein's Foresight to Bell's Oversight, and Back}
\vspace{-1.0cm}
\author{{\Large Joy Christian}\thanks{jjc@alum.bu.edu}
\\ {\small Wolfson College, University of Oxford,} \\ {\small Oxford, OX2 6UD, United Kingdom}}
\date{\begin{quote}
{\bf Abstract}:
It is well known that quantum correlations are not only more disciplined (and hence stronger) compared to classical
correlations, but they are more disciplined in a mathematically very precise sense. This raises an important physical
question: What is responsible for making quantum correlations so much more disciplined? Here we explain the observed
discipline of quantum correlations by identifying the symmetries of our physical space with those of a parallelized 7-sphere.
We substantiate this identification by proving that any quantum correlation can be understood as a classical, local-realistic
correlation among a set of points of a parallelized 7-sphere.\break
\end{quote}}
\maketitle
\end{titlepage}}
\chapter{On the Local-Realistic Origins of Quantum Correlations}
\pagenumbering{arabic}

\begin{center}
{\raggedleft In a good mystery story the most obvious} \\
${\!\!\!\!\!\!\!\!}${\raggedleft clues often lead to the wrong suspects.} \\
\smallskip
\smallskip
${\,\;\;\;\;\;\;\;\;\;\;\;\;\;\;\;\;\;\;\;\;\;\;\;\;\;\;\;\;\;\;\;\;\;\;\;\;}$Einstein and Infeld\\
\end{center}

{\baselineskip 11pt

\section{\large Introduction}

In 1927, when quantum theory was still in its infancy and John Bell was yet to be born, Albert Einstein---one of the founding
fathers of the theory---was attending the now famous ${5^{th}}$ Solvay Conference in Brussels. He was profoundly disturbed by what
the new theory had to say about the nature of physical reality. Among other things, his concerns stemmed from a deep appreciation
of unity in nature. Beyond the clich\'e of ``God does not play dice'', he had recognized that quantum theory entailed
a fundamental schism in nature. The aphorism ``God does not split reality'' perhaps better captures the true essence of his
concerns \cite{48-1}. What he sought was a unified picture of nature, devoid of any subjective boundary between the
classical and the quantum. What he suspected was a deeper layer of reality, beyond the polarized picture
offered by quantum theory\index{quantum theory}.

By 1935---when John Bell was seven---these worries of Einstein had matured into a powerful logical argument against the new
theory. Published in collaboration with Boris Podolsky and Nathan Rosen, this argument proves, {\it once and for all}, that
quantum theory provides at best an incomplete description of the physical reality \cite{EPR-1}.  Since the argument itself is
logically impeccable, this conclusion is beyond dispute. Any argument, however, can only be as good as its premises, and
that---as is well known---is where Bell entered the game in 1964 \cite{Bell-1964-1}. He attempted to show that not all of the
premises of EPR are mutually compatible. Ironically, however, it is the argument of Bell that turns out to contain a faulty
assumption, not that of EPR. What is more, this assumption appears in the very first equation of Bell's famous paper
\cite{Bell-1964-1}, and yet it had escaped notice until recently.

In a series of papers, written between 2007 and 2011, I tried to bring out Bell's error and constructed explicit counterexamples,
not only to his original theorem, but also to several of its variants. This book is a collection of these papers, each of which
can be read more or less independently, but their contents are interconnected, and\break
reveal different aspects of the fundamental
flaw in Bell's argument. The collection as a whole, however, is better viewed as addressing a very important physical question.
Regardless of the validity of his theorem, what Bell discovered in 1964 is physically quite significant. He discovered that
quantum correlations are far more disciplined than any classically possible correlation. What is more, quantum correlations
are not only more disciplined, but are more disciplined in a mathematically very precise
sense. This tells us something much more profound about the structure of the world we live in. And, at the same time,
it raises a very important physical question:
\vspace{-0.2cm}
\begin{center}
{\bf What is it that makes quantum correlations\break ${\,}$more disciplined than classical correlations?}
\end{center}
\vspace{-0.2cm}
My goal in this book is to answer this question in mathematically and physically as precise a sense as possible.
To this end, let me begin with an extended summary of my argument against Bell's theorem.

As noted above, the story began with Einstein, Podolsky, and Rosen${\,}$\cite{EPR-1}.
The logic of their argument can be summarized as follows:
\vspace{-1.2cm}
\begin{center}
\item[] ${\,\;\;\;\;\;\;}${\bf (1)} ${\;}$QM ${\!\implies\!}$ Perfect Correlations
\item[] ${\;\;\;\;\;\;\;\;\;+\,\;\;}${\bf(2)} ${\;}$Adherence to Local Causality${\,\;\;\;\;\;\;\;}$
\item[] ${\,\;\;\;\;\;\;+\;\;\;}${\bf(3)} ${\;}$Criterion of Objective Reality${\;\;\;\;\,}$
\item[] ${\;\;+\;\;\;}${\bf(4)} ${\;}$Notion of a Complete Theory${\,}$
\item[] ${\;\!\implies\!\!}$ {\bf (5)} ${\;}$QM is an Incomplete Theory.${\;}$
\end{center}
\vspace{-0.1cm}
Given their premises, the conclusion of EPR follows impeccably. Among their premises (which are hardly
unreasonable), the one that concerns us the most is their criterion of completeness:
\vspace{-0.2cm}
\begin{center}
{\bf every element of the physical reality must\break have a counterpart in the physical theory.}
\end{center}
\vspace{-0.2cm}
Bell attempted to prove that no theory satisfying this criterion can be locally causal. To this end,
he took a complete theory to mean any theory whose predictions are dictated by functions of the form
\begin{equation}
{\mathscr A}({\bf n},\,\lambda): {\rm I\!R}^3\!\times\Lambda\longrightarrow S^0\equiv \{-1,\,+1\}, \label{map-1-1}
\end{equation}
where ${{\rm I\!R}^3}$ is the space of 3-vectors, ${\Lambda}$ is a space of ``complete'' states, and ${S^0\equiv \{-1,\,+1\}}$
is a unit 0-sphere. He then claimed (correctly, as it turns out)
that no pair of functions of this form can reproduce the correlations for the singlet
state predicted by quantum mechanics\footnote{This is hardly surprising. After all, the product moment
correlation coefficient employed by Bell in his paper, by
definition, is a measure of {\it linear} ${\,}$relationship between bivariate variables. Thus Bell implicitly assumed a
linear relationship between ${\mathscr A}$ and ${\mathscr B}$ to prove that the relationship between them must be linear!}:
\begin{equation}
\langle\, {\mathscr A}({\bf a},\,\lambda)\,{\mathscr B}({\bf b},\,\lambda)\, \rangle\,\not=\,
-\,{\bf a}\cdot{\bf b}\,. \label{corre-1-1}
\end{equation}

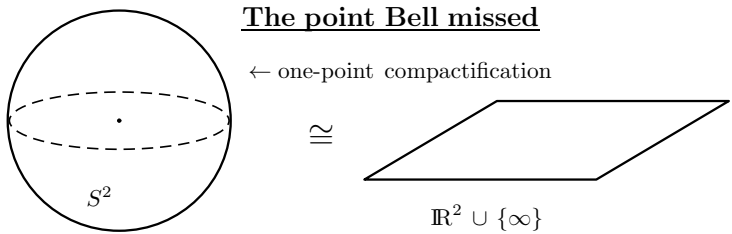
\begin{figure}
\hrule
\vspace{-0.45cm}
\scalebox{0.87}{
\begin{pspicture}(3.2,-1.9)(-0.27,3.0)

\pscircle[linewidth=0.35mm](1.8,0.0){1.7}

\psline[linewidth=0.05mm,dotsize=1.5pt 3]{*-}(1.8,0.0)(1.805,0)

\psellipse[linewidth=0.25mm,linestyle=dashed](1.8,0.0)(1.67,0.45)

\pspolygon[linewidth=0.35mm](5.5,-0.9)(9.0,-0.9)(11.0,0.3)(7.5,0.3)

\put(1.3,-1.3){${S^2}$}

\put(3.65,+1.45){\large{\underbar{\bf The\;point\;Bell\;missed}}}

\put(3.75,0.65){${\leftarrow\;}${\rm one-point\;\,compactification}}

\put(6.5,-1.6){${{\rm I\!R}^2\,\cup\,\{\infty\}}$}

\put(4.65,-0.4){\Large ${\cong}$}

\end{pspicture}}
\vspace{0.2cm}
\hrule
\caption{Although lines and planes contain the same number of points, it is impossible to put the points of a line or a plane
in a one-to-one correspondence with all of the points of a 2-sphere.\break}
\label{fig-11}
\hrule
\end{figure}

At first sight, this appears to be a straightforward mathematical contradiction undermining the force of the EPR
argument \cite{Bell-1964-1}. And for this reason functions of this form are routinely assumed in the Bell literature
to provide complete specifications of the elements of physical reality, or complete accounting of all possible measurement results.
As we shall see however, Bell's prescription is not only false, it is breathtakingly na\"ive and unphysical.
It stems from an incorrect underpinning of both the EPR argument and
the actual topological configurations involved in the relevant experiments \cite{What-1}. In truth, for any two-level system
the EPR criterion of completeness demands that the correct functions must necessarily be of the form
\begin{equation}
\pm\,1\,=\,{\mathscr A}({\bf n},\,\lambda):
{\rm I\!R}^3\!\times\Lambda\longrightarrow S^3 \hookrightarrow {\rm I\!R}^4, \label{map-1-2}
\end{equation}
with the {\it simply-connected} codomain ${S^3}$ of ${{\mathscr A}({\bf n},\,\lambda)}$ replacing the
{\it totally disconnected} codomain ${S^0}$ assumed by Bell.
It is important to note here that this correction does not affect the actual
measurement results. For a specific vector ${\bf n}$ and an initial state ${\lambda}$ we still have
\begin{equation}
{\mathscr A}({\bf n},\,\lambda)\,=\,+\,1\;\,{\rm or}\;-1
\end{equation}
as demanded by Bell, but now the topology of the codomain of the function ${{\mathscr A}({\bf n},\,\lambda)}$ has changed
from a 0-sphere to a 3-sphere, with the latter embedded in ${{\rm I\!R}^4}$ in such a manner that the above constraint is
satisfied. On the other hand, as is evident from Fig.${\,}$(\ref{fig-11}) (and will be further clarified in the following
pages), without this topological correction it is impossible to provide a complete account of all possible measurement
results. Thus the selection of the codomain ${S^3\hookrightarrow {\rm I\!R}^4}$
in equation (\ref{map-1-2}) is not a matter of choice but necessity.
What is responsible for the EPR correlations is {\it the topology of the set of all possible measurement results} \cite{What-1}.
And for a two-level system this set happens to be an equatorial 2-sphere within a parallelized 3-sphere. But once the codomain of
the functions ${{\mathscr A}({\bf n},\,\lambda)}$
is so corrected, the proof of Bell's theorem (as given in Ref.${\,}$\cite{Bell-1964-1}) simply falls apart. In fact, as we shall
repeatedly see in the following pages, the strength of the correlation for {\it any}${\,}$ physical system is entirely determined
by the topology of the codomain of the local functions ${{\mathscr A}({\bf n},\,\lambda)}$. It has nothing whatsoever to do with
entanglement or nonlocality.}

{\baselineskip 11.41pt

\section{\large Local Origins of the EPR-Bohm Correlations}

Put differently, once the measurement results are represented by functions of the form (\ref{map-1-2}),
it is quite easy to reproduce
the quantum correlations purely classically, in a manifestly local-realistic manner. For example, suppose Alice and Bob are equipped
with the variables
\begin{align}
{\mathscr A}({\bf a},\,{\lambda})\,=\,\{-\,a_j\;{\boldsymbol\beta}_j\,\}\,\{\,a_k\;{\boldsymbol\beta}_k(\lambda)\,\}\,&=\,
\begin{cases}
+\,1\;\;\;\;\;{\rm if} &\lambda\,=\,+\,1 \\
-\,1\;\;\;\;\;{\rm if} &\lambda\,=\,-\,1
\end{cases} \label{comeout-1}
\end{align}
and
\begin{align}
{\mathscr B}({\bf b},\,{\lambda})\,=\,\{+\,b_k\;{\boldsymbol\beta}_k\,\}\,\{\,b_j\;{\boldsymbol\beta}_j(\lambda)\,\}\,&=\,
\begin{cases}
-\,1\;\;\;\;\;{\rm if} &\lambda\,=\,+\,1 \\
+\,1\;\;\;\;\;{\rm if} &\lambda\,=\,-\,1\,,
\end{cases} \label{comeout-2}
\end{align}
where the repeated indices are summed over ${x,\,y,}$ and ${z}$; the fixed bivector
basis ${\{{\boldsymbol\beta}_x,\,{\boldsymbol\beta}_y,\,{\boldsymbol\beta}_z\}}$ is defined by the
algebra
\begin{equation}
{\boldsymbol\beta}_j\,{\boldsymbol\beta}_k \,=\,-\,\delta_{jk}\,-\,\epsilon_{jkl}\,{\boldsymbol\beta}_l\,;\label{of-which}
\end{equation}
and---together with ${{\boldsymbol\beta}_j(\lambda)=\lambda\,{\boldsymbol\beta}_j}$---the ${\lambda}$-dependent bivector basis
${\{{\boldsymbol\beta}_x(\lambda),\,{\boldsymbol\beta}_y(\lambda),\,{\boldsymbol\beta}_z(\lambda)\}}$ is defined by the algebra
\begin{equation}
{\boldsymbol\beta}_j\,{\boldsymbol\beta}_k\,=\,-\,\delta_{jk}\,-\,\lambda\;\epsilon_{jkl}\,
{\boldsymbol\beta}_l\,,\label{nof-which}
\end{equation}
where ${\lambda\,=\,\pm\,1}$ is a fair coin representing two alternative orientations of the 3-sphere\footnote{Needless to say,
${{\mathscr A}({\bf a},\,{\lambda})}$ and ${{\mathscr B}({\bf b},\,{\lambda})}$ are two
{\it different} functions of the random variable ${\lambda}$. Moreover, they are {\it statistically independent events}
occurring within a 3-sphere, with factorized joint probability
${P({\mathscr A}\;\text{and}\;{\mathscr B})=P({\mathscr A})\times P({\mathscr B})\leq\frac{1}{2}}$.
Therefore their product ${{\mathscr A}{\mathscr B}}$ is guaranteed to be equal to ${-1}$ only for the
case ${{\bf a}={\bf b}}$. For all other ${\bf a}$ and ${\bf b}$, ${{\mathscr A}{\mathscr B}}$ will
alternate between the values ${-1\;\text{and}\;+1}$.},
${\delta_{jk}}$ is the Kronecker delta, ${\epsilon_{jkl}}$ is the Levi-Civita symbol,
and ${{\bf a}=a_j\,{\bf e}_j}$ and ${{\bf b}=b_j\,{\bf e}_j}$ are unit vectors \cite{Clifford-1}. Evidently, the variables
${{\mathscr A}({\bf a},\,{\lambda})}$ and ${{\mathscr B}({\bf b},\,{\lambda})}$ belonging to ${S^3}$---in addition to being
manifestly realistic---are strictly {\it local} variables. In fact, they are not even contextual \cite{Contextual-1}.
Alice's measurement result---although it refers to a freely chosen direction ${\bf a}$---depends only on the initial state
${\lambda}$; and likewise, Bob's measurement result---although it refers to a freely chosen direction ${\bf b}$---depends
only on the initial state ${\lambda}$.
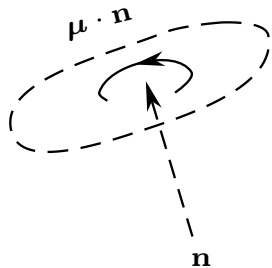
\begin{figure}
\hrule
\vspace{0.2cm}
\scalebox{1.53}{
\begin{pspicture}(-0.7,-2.3)(-6.9,-4.95)

\begin{rotate}{-180}

\pscurve[linewidth=0.2mm,linestyle=dashed](3.25,2.64)(3.75,2.73)(5.05,3.6)(4.0,3.6)(2.85,2.75)(3.25,2.64)

\rput{180}(3.425,4.7){\scriptsize {${\bf n}$}}

\psline[linewidth=0.2mm,linestyle=dashed,arrowinset=0.3,arrowsize=2pt 3,arrowlength=2]{<-}(3.9,3.15)(3.5,4.5)

\pscurve[linewidth=0.2mm]{-}(3.65,3.25)(3.51,3.07)(4.3,3.25)(4.23,3.32)

\psline[linewidth=0.2mm,arrowinset=0.3,arrowsize=2pt 3,arrowlength=2]{->}(3.9,2.985)(4.0,3.005)

\rput{194}(4.3,2.65){\scriptsize {${{\boldsymbol\mu}\cdot{\bf n}}$}}

\end{rotate}

\end{pspicture}}
\vspace{0.2cm}
\hrule
\caption{A unit bivector represents an equatorial point of a unit, parallelized 3-sphere. It is
an abstraction of a directed plane segment,\break with only a magnitude and a sense of rotation---{\it i.e.}, 
clockwise (${-}$) or counterclockwise (${+}$). Neither the depicted oval shape of its plane,
nor its axis of rotation ${\bf n}$, is an intrinsic part of the bivector ${{\boldsymbol\mu}\cdot{\bf n}}$.\break}
\label{fig-44}
\hrule
\end{figure}

In the subsequent chapters we shall mainly use the standard notations of Clifford algebra ${{Cl}_{3,0}}$. The bivector algebras
(\ref{of-which}) and (\ref{nof-which})
will then be seen as even subalgebras of ${{Cl}_{3,0}}$. The latter is a linear vector space, ${{\rm I\!R}^8}$,
spanned by the graded orthonormal basis
\begin{equation}
\left\{1,\,\;{\bf e}_x,\,{\bf e}_y,\,{\bf e}_z,\,\;{\bf e}_x\wedge{\bf e}_y,\,
{\bf e}_y\wedge{\bf e}_z,\,{\bf e}_z\wedge{\bf e}_x,\,\;
{\bf e}_x\wedge{\bf e}_y\wedge{\bf e}_z\right\}\!,
\end{equation}
where ``${\wedge}$'' is the outer product, and the trivector
${I\equiv{\bf e}_x\wedge{\bf e}_y\wedge{\bf e}_z}$ defines the fundamental volume form of the physical space. In terms of these
notations we can rewrite the bivector ${\{\,a_j\;{\boldsymbol\beta}_j(\lambda)\,\}}$ as
\begin{equation}
{\boldsymbol\mu}\cdot{\bf a}\,\equiv\,\{\,a_j\;{\boldsymbol\beta}_j(\lambda)\,\}\,\equiv\,\lambda\,
\{\,a_x\;{{\bf e}_y}\,\wedge\,{{\bf e}_z}
\,+\,a_y\;{{\bf e}_z}\,\wedge\,{{\bf e}_x}
\,+\,a_z\;{{\bf e}_x}\,\wedge\,{{\bf e}_y}\},
\end{equation}
with ${{\boldsymbol\mu}=\lambda\,I}$ now representing the hidden variable of the model. The variables
${{\mathscr A}({\bf a},\,{\lambda})}$ and ${{\mathscr B}({\bf b},\,{\lambda})}$ defined above then take the form
\begin{equation}
S^3\ni {\mathscr A}({\bf a},\,{\boldsymbol\mu})\,=\,(-\,I\cdot{{\bf a}}\,)
\,(\,+\,{\boldsymbol\mu}\cdot{{\bf a}}\,)\,=\,
\begin{cases}
+\,1\;\;\;\;\;{\rm if} &{\boldsymbol\mu}\,=\,+\,I \\
-\,1\;\;\;\;\;{\rm if} &{\boldsymbol\mu}\,=\,-\,I
\end{cases} \label{17-noy-111}
\end{equation}
and
\begin{equation}
S^3\ni {\mathscr B}({\bf b},\,{\boldsymbol\mu})\,=\,(+\,I\cdot{{\bf b}}\,)
\,(\,+\,{\boldsymbol\mu}\cdot{{\bf b}}\,)\,=\,
\begin{cases}
-\,1\;\;\;\;\,{\rm if} &{\boldsymbol\mu}\,=\,+\,I \\
+\,1\;\;\;\;\,{\rm if} &{\boldsymbol\mu}\,=\,-\,I,
\end{cases} \label{18-noy-111}                
\end{equation}
with the trivector ${\boldsymbol\mu}$ being either ${+\,I}$ or ${-\,I}$ with equal probability. In what follows we shall
view the fixed bivectors ${(-\,I\cdot{{\bf a}}\,)}$ and ${(+\,I\cdot{{\bf b}}\,)}$ as representing the measuring instruments
for detecting the random bivectors ${(\,+\,{\boldsymbol\mu}\cdot{{\bf a}}\,)}$ and
${(\,+\,{\boldsymbol\mu}\cdot{{\bf b}}\,)}$, which represent the spins.

It is crucial to note that the variables ${{\mathscr A}({\bf a},\,{\lambda})}$ and ${{\mathscr B}({\bf b},\,{\lambda})}$
are generated with {\it different} bivectorial scales of dispersion (or different standard deviations)
for each direction ${\bf a}$ and  ${\bf b}$. Consequently,
in statistical terms these variables are raw scores, as opposed to standard scores \cite{scores-1}.
Recall that a standard score indicates how
many standard deviations an observation or datum is above or below the mean. If ${\rm x}$ is a raw (or unnormalized) score
and ${\overline{\rm x}}$ is its mean value, then the standard (or normalized) score, ${{\rm z}({\rm x})}$, is defined by
\begin{equation}
{\rm z}({\rm x})\,=\,\frac{{\rm x}\,-\,{\overline{\rm x}}}{\sigma({\rm x})}\,,
\end{equation}
where ${\sigma({\rm x})}$ is the standard deviation of ${\rm x}$. A standard score thus represents the distance between
a raw score and the population mean in the units of standard deviation, and
allows us to make comparisons of raw scores that may have come from very different sources. In other words, the mean value of the
standard score itself is always zero, with standard deviation unity.
In terms of these concepts the bivariate correlation between raw scores ${\rm x}$ and ${\rm y}$ is defined as
\begin{align}
{\cal E}({\rm x},\,{\rm y})\,&=\;\frac{\,{\displaystyle\lim_{\,n\,\gg\,1}}\left[{\displaystyle\frac{1}{n}}\,
{\displaystyle\sum_{i\,=\,1}^{n}}\,({\rm x}^i\,-\,{\overline{\rm x}}\,)\;
({\rm y}^i\,-\,{\overline{\rm y}}\,)\right]}{\sigma({\rm x})\;\sigma({\rm y})} \label{co} \\
&=\,\lim_{\,n\,\gg\,1}\left[\frac{1}{n}
\sum_{i\,=\,1}^{n}\,{\rm z}({\rm x}^i)\;{\rm z}({\rm y}^i)\right]. \label{stan}
\end{align}
It is vital to appreciate that covariance by itself---{\it i.e.}, the numerator of equation (\ref{co}) by itself---does not
provide the correct measure of association between the raw scores, not the least because it depends on different units and scales
(or different scales of dispersion) that may have been used (advertently or inadvertently) in the measurements of such scores
\cite{scores-1}. Therefore, to arrive at the correct measure of association between the raw scores one must either use equation
(\ref{co}), with the product of standard
deviations in the denominator, or use covariance of the standardized variables, as in Eq.${\,}$(\ref{stan}).

These basic statistical concepts are crucial for understanding the EPR correlations.
As defined above, the random variables ${{\mathscr A}({\bf a},\,{\lambda})}$ and ${{\mathscr B}({\bf b},\,{\lambda})}$ are
products of two factors---one random and another non-random. Within ${{\mathscr A}({\bf a},\,{\lambda})}$ the factor
${\{\,a_k\;{\boldsymbol\beta}_k(\lambda)\,\}}$ is a random factor---a function of the hidden variable ${\lambda}$, whereas
${\{-\,a_j\;{\boldsymbol\beta}_j\,\}}$ is a non-random factor, independent of the hidden variable ${\lambda}$. Thus,
as a random variable each number ${{\mathscr A}({\bf a},\,{\lambda})}$ and ${{\mathscr B}({\bf b},\,{\lambda})}$ is
generated with a {\it different} standard deviation---{\it i.e.}, a {\it different} size of typical error.
More specifically, ${{\mathscr A}({\bf a},\,{\lambda})}$ is generated with the standard deviation
${\{-\,a_j\;{\boldsymbol\beta}_j\,\}}$, whereas ${{\mathscr B}({\bf b},\,{\lambda})}$ is generated
with a different standard deviation, namely ${\{+\,b_k\;{\boldsymbol\beta}_k\,\}}$. These two deviations
can be calculated easily. Since errors in linear relations propagate linearly,
the standard deviation of ${{\mathscr A}({\bf a},\,{\lambda})}$ is equal to ${\{-\,a_j\;{\boldsymbol\beta}_j\,\}}$
times the standard deviation of ${\{\,a_k\;{\boldsymbol\beta}_k(\lambda)\,\}}$ (which we write as ${\sigma({A})}$), whereas
the standard deviation of ${{\mathscr B}({\bf b},\,{\lambda})}$ is equal to ${\{+\,b_k\;{\boldsymbol\beta}_k\,\}}$
times the standard deviation of ${\{\,b_j\;{\boldsymbol\beta}_j(\lambda)\,\}}$ (which we write as ${\sigma({B})}$):
\begin{align}
\sigma({\mathscr A}\,)\,&=\,\{-\,a_j\;{\boldsymbol\beta}_j\,\}\,\sigma({A}) \notag \\
\text{and}\;\;\sigma({\mathscr B}\,)\,&=\,\{+\,b_k\;{\boldsymbol\beta}_k\,\}\,\sigma({B}).
\end{align}
But since all the bivectors we have been considering are normalized to unity, and since the mean of
${\{\,a_k\;{\boldsymbol\beta}_k(\lambda)\,\}}$ vanishes on the account of ${\lambda}$ being a
fair coin, its standard deviation is easy to calculate, and it turns out to be equal to unity:
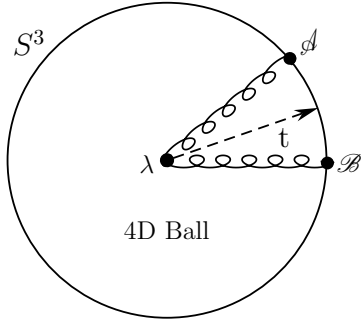
\begin{figure}
\hrule
\scalebox{0.81}{
\begin{pspicture}(-8.0,-3.7)(4.2,2.8)

\pscircle[linewidth=0.3mm](-1.8,-0.45){2.6}

\psarc[linewidth=0.0pt,linestyle=none,dotsize=6pt 3]{*-*}(-1.8,-0.45){2.6}{359}{40}

\psline[linewidth=0.3mm,linestyle=dashed,dotsize=4pt 3,arrowinset=0.3,arrowsize=3pt 3,arrowlength=2]{*->}(-1.8,-0.45)(0.65365,0.41)

\pscoil[coilarmA=-0.001,coilarmB=0.00001,coilwidth=0.2,coilheight=3,coilaspect=315]{}(0.16,1.2)(-1.745,-0.39)

\pscoil[coilarmA=-0.001,coilarmB=-0.00001,coilwidth=0.2,coilheight=3,coilaspect=315]{}(-1.74,-0.47)(0.69,-0.47)

\put(-4.3,1.27){{\Large ${S^3}$}}

\put(-2.5,-1.75){\large{4D Ball}}

\put(-0.0,-0.23){\Large{t}}

\put(-2.25,-0.65){\large{${\lambda}$}}

\rput{25}(0.5,1.48){\large{${\mathscr A}$}}

\put(0.95,-0.64){\large{${\mathscr B}$}}

\end{pspicture}}
\hrule
\caption{An initial EPR state ${\lambda}$ originated at time ${{\rm t}=0}$ evolves into measurement results ${\mathscr A}$ and
${\mathscr B}$ at a later time, occurring at two spacelike separated locations on
a parallelized 3-sphere, ${S^3}$, which can be thought of as a boundary of a 4-dimensional ball of radius t. In what
follows we shall assume that t has been normalized to unity.\break}
\label{fig-33}
\hrule
\end{figure}
\begin{align}
\sigma({A})\,&=\,\sqrt{\frac{1}{n}\sum_{i\,=\,1}^{n}\,\left|\left|\,A({\bf a},\,{\lambda}^i)\,-\,
{\overline{A({\bf a},\,{\lambda}^i)}}\;\right|\right|^2\,}\, \notag \\
&=\,\sqrt{\frac{1}{n}\sum_{i\,=\,1}^{n}\,
\left|\left|\,\{\,a_k\;{\boldsymbol\beta}_k(\lambda^i)\,\}\,-\,0\,\right|\right|^2\,}\,=\,1,
\end{align}
where the last equality follows from the fact that ${\{\,a_k\;{\boldsymbol\beta}_k(\lambda^i)\,\}}$ are normalized to unity.
Similarly, we find that ${\sigma({B})}$ is also equal to ${1}$. As a result, the standard deviation of
${{\mathscr A}({\bf a},\,{\lambda})}$ works out to be equal to ${\{-\,a_j\;{\boldsymbol\beta}_j\,\}}$, and the standard
deviation of ${{\mathscr B}({\bf b},\,{\lambda})}$ works out to be equal to ${\{+\,b_k\;{\boldsymbol\beta}_k\,\}}$. Putting
these two results together, we arrive at the following standardized scores corresponding to the raw scores:
\begin{align}
A({\bf a},\,{\lambda})&=\frac{\,{\mathscr A}({\bf a},\,{\lambda})\,-\,
{\overline{{\mathscr A}({\bf a},\,{\lambda})}}}{\sigma({\mathscr A})} \notag \\
\,&=\,\frac{\,{\mathscr A}({\bf a},\,{\lambda})\,-\,0\,}{\{-\,a_j\;{\boldsymbol\beta}_j\,\}}
\,=\,\{\,a_k\;{\boldsymbol\beta}_k(\lambda)\,\} \label{var-a}
\end{align}
\begin{align}
\!\!\!\!\!\!\!\!\!\!\!\!\!\!\!\!\!\!\!\!
\text{and}\;\;\;\;\;\;\;\;\;\;\;\;\;\;\;
B({\bf b},\,{\lambda})&=\frac{\,{\mathscr B}({\bf b},\,{\lambda})\,-\,
{\overline{{\mathscr B}({\bf b},\,{\lambda})}}}{\sigma({\mathscr B})} \notag \\
\,&=\,\frac{\,{\mathscr B}({\bf b},\,{\lambda})\,-\,0\,}{\{+\,b_k\;{\boldsymbol\beta}_k\,\}}
\,=\,\{\,b_j\;{\boldsymbol\beta}_j(\lambda)\,\},\label{var-b}
\end{align}
where we have used the identities such as ${\{+\,a_k\;{\boldsymbol\beta}_k\,\}\{-\,a_k\;{\boldsymbol\beta}_k\,\}=+1}$.
Not surprisingly, just like the raw scores ${{\mathscr A}({\bf a},\,{\lambda})}$ and ${{\mathscr B}({\bf b},\,{\lambda})}$,
these standard scores are also strictly {\it local} variables: ${A({\bf a},\,{\lambda})}$ depends only on the freely chosen
local direction ${\bf a}$ and the common cause ${\lambda}$, and likewise ${B({\bf b},\,{\lambda})}$ depends only on the
freely chosen local direction ${\bf b}$ and the common cause ${{\lambda}\,}$.
Moreover, despite appearances, ${\,A({\bf a},\,{\lambda})}$ and ${B({\bf b},\,{\lambda})\,}$ are simply binary
numbers, ${\pm\,1}$, albeit occurring within the compact topology of the 3-sphere rather than the real line:
\begin{align}
S^3\supset S^2\ni A({\bf a},\,{\lambda})&\,=\,\{\,a_k\;{\boldsymbol\beta}_k(\lambda)\,\}
\,=\, \pm\,1\;\,{\rm about}\;{{\bf a}}\in{\rm I\!R}^3, \\
S^3\supset S^2\ni B({\bf b},\,{\lambda})&\,=\,\,\{\,b_j\;{\boldsymbol\beta}_j(\lambda)\,\}
\,=\, \pm\,1\;\,{\rm about}\;{{\bf b}}\in{\rm I\!R}^3.
\end{align}
In fact, since the space of all bivectors ${\{\,a_k\;{\boldsymbol\beta}_k(\lambda)\,\}}$ is isomorphic to the equatorial 2-sphere
contained within the 3-sphere \cite{What-1}, each standard score ${A({\bf a},\,{\lambda})}$ of Alice is uniquely identified with
a definite point of this 2-sphere, and likewise for the standard scores of Bob.

\vspace{5pt}

{\parbox{9cm}{\bf In the following chapters we shall tacitly assume that this procedure of standardizing from the raw
scores to standard scores has been performed for all measurement results, taken either as points of a 3-sphere, or more
generally as points of a 7-sphere.}}

\vspace{5pt}

Now, since we have assumed that initially there was 50/50 chance between the right-handed and left-handed orientations
of the 3-sphere ({\it i.e.}, equal chance between the initial states ${{\lambda}=+\,1}$ and ${{\lambda}=-\,1}$), the
expectation values of the local outcomes vanish trivially. On the other hand, as discussed above, to determine the
{\it correct} correlation between the joint observations of Alice and Bob we must calculate covariance between the
corresponding standard scores ${{A}({\bf a},\,{\lambda})}$ and ${{B}({\bf b},\,{\lambda})}$, not the raw scores themselves.
The correlation between the raw scores ${{\mathscr A}({\bf a},\,{\lambda})}$ and ${{\mathscr B}({\bf b},\,{\lambda})}$
then works out to be
\begin{align}
{\cal E}({\bf a},\,{\bf b})\,&=\,\lim_{\,n\,\gg\,1}\left[\frac{1}{n}\sum_{i\,=\,1}^{n}\,A({\bf a},\,{\lambda}^i)\,
B({\bf b},\,{\lambda}^i)\right] \label{55} \\
&=\lim_{\,n\,\gg\,1}\left[\frac{1}{n}\sum_{i\,=\,1}^{n}\,
\left\{\,a_j\;{\boldsymbol\beta}_j(\lambda^i)\,\right\}\,\left\{\,b_k\;{\boldsymbol\beta}_k(\lambda^i)\,\right\}\right]  \\
&=\,-\,a_j\,b_j\,-\lim_{\,n\,\gg\,1}\left[\frac{1}{n}\sum_{i\,=\,1}^{n}\,
\left\{\,\lambda^i\,\epsilon_{jkl}\;a_j\,b_k\;{\boldsymbol\beta}_l\,\right\}\right] \\
&=\,-\,a_j\,b_j\,+\,0\,=\,-\,{\bf a}\cdot{\bf b}\,,
\end{align}
where we have used the algebra defined in (\ref{nof-which}). Consequently, when the raw scores ${{\mathscr A}=\pm\,1}$ and
${{\mathscr B}=\pm\,1}$ are compared in practice by coincidence counts \cite{Aspect-1}\cite{Weihs-1}, the normalized
expectation value of their product will inevitably yield
\begin{align}
{\cal E}({\bf a},\,{\bf b})\,&=\,\frac{\Big[C_{++}({\bf a},\,{\bf b})\,+\,C_{--}({\bf a},\,{\bf b})
\,-\,C_{+-}({\bf a},\,{\bf b})\,-\,C_{-+}({\bf a},\,{\bf b})\Big]}{\Big[C_{++}({\bf a},\,{\bf b})\,+\,C_{--}({\bf a},\,{\bf b})
\,+\,C_{+-}({\bf a},\,{\bf b})\,+\,C_{-+}({\bf a},\,{\bf b})\Big]}\, \notag \\
&=\,-\,{\bf a}\cdot{\bf b}\,, \label{calcul}
\end{align}
where ${C_{+-}({\bf a},\,{\bf b})}$ {\it etc.}${\;}$represent the number of joint occurrences of detections ${+\,1}$ along ${\bf a}$
and ${-\,1}$ along ${\bf b}$ {\it etc}.

\begin{figure}
\hrule
\scalebox{0.81}{
\begin{pspicture}(-3.9,-1.0)(4.87,5.5)
\psset{xunit=0.5mm,yunit=4cm}
\psaxes[axesstyle=frame,tickstyle=top,dx=90\psxunit,Dx=180,dy=1\psyunit,Dy=+2,Oy=-1](0,0)(90,1.0)
\psplot[linewidth=0.33mm]{0.0}{90}{x dup cos exch cos mul 1.0 mul neg 1 add}
\psline[linewidth=0.2mm,arrowinset=0.3,arrowsize=2pt 3,arrowlength=2]{->}(0,0.5)(100,0.5)
\psline[linewidth=0.2mm]{-}(45,0)(45,1)
\psline[linewidth=0.25mm,arrowinset=0.3,arrowsize=2pt 3,arrowlength=2]{->}(0,1)(0,1.2)
\psline[linewidth=0.33mm,linestyle=dashed]{-}(0,0)(90,1)
\put(2.1,-0.54){${90}$}
\put(-0.75,3.92){${+}$}
\put(-0.4,4.7){\large ${\cal E}$}
\put(-0.38,1.93){${0}$}
\put(5.13,1.85){\large ${\theta}$}
\end{pspicture}}
\hrule
\caption{Local-realistic correlations can be stronger within ${S^3}$.\break\break}
\vspace{0.25cm}
\label{fig-55}
\hrule
\end{figure}
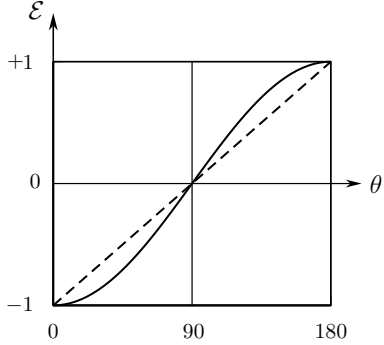

The above equation simply describes covariance of the numbers ${{\mathscr A}=\pm\,1}$ and ${{\mathscr B}=\pm\,1}$ yielding 
the ``impossible'' strong correlation:
\begin{align}
{\cal E}({\bf a},\,{\bf b})\,=\lim_{\,n\,\gg\,1}\left[\frac{1}{n}\sum_{i\,=\,1}^{n}\,
{\mathscr A}({\bf a},\,{\lambda}^i)\;{\mathscr B}({\bf b},\,{\lambda}^i)\right]
\,=\,-\,{\bf a}\cdot{\bf b}\,.\label{qcorraw}
\end{align}
How is this ``impossible'' result possible? After all, Bell seems to have proved mathematically \cite{Bell-1964-1} that correlation
between such numbers can never exceed the linear limit. The answer is that, apart from the statistical differences we discussed
above, there are also topological differences between the above expression and what Bell considered in his theorem. In particular,
in the above equation the numbers ${{\mathscr A}=\pm\,1}$ and ${{\mathscr B}=\pm\,1}$ are points of a parallelized 3-sphere,
\begin{equation}
\pm\,1\,=\,{\mathscr A}
({\bf n},\,\lambda): {\rm I\!R}^3\!\times\Lambda\longrightarrow S^3 \hookrightarrow{\rm I\!R}^4, \label{map-456}
\end{equation}
and since a parallelized 3-sphere is a compact, simply-connected topological space, it should not be surprising that
correlation among its points is stronger-than-linear. By contrast in Bell's argument the measurement functions
are implicitly assumed to be of the form
\begin{equation}
\pm\,1\,=\,{\mathscr A}({\bf n},\,\lambda): {\rm I\!R}^3\!\times\Lambda\longrightarrow {\cal I}\subseteq{\rm I\!R}\,,\label{map-121}
\end{equation}
and since the topology of the real line is far less disciplined than that of a 3-sphere, it is impossible to generate strong
correlation among its points. In other words, the strength of the correlation is entirely dependent on how disciplined the
codomain of the functions ${{\mathscr A}({\bf n},\,\lambda)}$ is. Now it turns out that the parallelized spheres ${S^3}$ and
${S^7}$ have maximally disciplined topology in this respect \cite{What-1}.
And since the 3-sphere can be parallelized by unit quaternions, the
covariance between its equatorial points represented by pure quaternions (or unit bivectors) is precisely the EPR correlation
\begin{align}
{\cal E}({\bf a},\,{\bf b})\,=\lim_{\,n\,\gg\,1}\left[\frac{1}{n}\sum_{i\,=\,1}^{n}\,
{A}({\bf a},\,{\lambda}^i)\;{B}({\bf b},\,{\lambda}^i)\right]
\,=\,-\,{\bf a}\cdot{\bf b}\,,
\end{align}
where the bivectors ${{A}({\bf a},\,{\lambda})=\{\,a_k\;{\boldsymbol\beta}_k(\lambda)\}}$ and
${{B}({\bf b},\,{\lambda})=\{\,b_k\;{\boldsymbol\beta}_k(\lambda)\}}$ are the standardized variables
in the statistical terms discussed above.
Then, for arbitrary four directions ${\bf a}$, ${\bf a'}$, ${\bf b}$, and ${\bf b'}$,
the corresponding CHSH string of expectation values immediately gives
\begin{align}
|{\cal E}({\bf a},\,{\bf b})\,+\,{\cal E}({\bf a},\,{\bf b'})\,&+\,
{\cal E}({\bf a'},\,{\bf b})\,-\,{\cal E}({\bf a'},\,{\bf b'})|\, \notag \\
&\leq\,2\,\sqrt{\,1-({\bf a}\times{\bf a'})\cdot({\bf b'}\times{\bf b})}\,\leq\,2\sqrt{2}\,,
\end{align}
which squarely contradicts Bell's theorem and exactly reproduces the quantum mechanical prediction.
So much for Bell's theorem.

\section{\large Local Origins of ALL Quantum Correlations}

What the above example shows is that the correlation among the raw scores observed by Alice and Bob are entirely
determined by the topology of the codomain of the local functions ${{\mathscr A}({\bf a},\,{\lambda})}$ and
${{\mathscr B}({\bf b},\,{\lambda})}$. In particular, correlations among the points of a unit parallelized
3-sphere are stronger than those among the points of the real line. Moreover,
as we shall see in the following chapters, once the topology
of the codomain is correctly specified, not only the EPR correlations, but also the correlations predicted by
the rotationally non-invariant quantum states---such as the GHZ states and Hardy state---can be exactly reproduced
in a purely local-realistic manner. Thus, contrary to the widespread belief, the correlations exhibited by such
states are not some irreducible quantum effects, but purely local-realistic, topological effects. In the cases of
EPR and Hardy states the correct topology of the codomain is a parallelized 3-sphere, and consequently
the correlations exhibited by these states are classical correlations among the points of a parallelized 3-sphere.
In the case of GHZ states on the other hand the correct topology of the codomain is a parallelized
7-sphere, and consequently the correlations exhibited by these states are classical correlations among the points of a
parallelized\break 7-sphere. More generally, all quantum mechanical correlations can be understood as purely classical,
local-realistic correlations among the points of a parallelized 7-sphere. Needless to say, this vindicates\break Einstein's
suspicion that a quantum state merely describes a statistical ensemble of physical systems, not an individual physical system.

What is more, as we shall see in Chapter 7 in greater detail, there are profound mathematical and conceptual reasons why the
topology of the 7-sphere plays such a significant role in the manifestation of {\it all} ${\,}$quantum correlations. As is well
known, quantum correlations are more disciplined (or stronger) than classical correlations in a mathematically precise sense.
It turns out that it is the discipline of parallelization in the manifold of all possible measurement results that is
responsible for the strength of quantum correlations. In fact, both the existence as well as the strength of all quantum
correlations are dictated by the parallelizability of the spheres ${S^0}$, ${S^1}$, ${S^3}$, and ${S^7}$, with the 7-sphere
being homeomorphic to the most general possible\break division algebra. And it is the property of division that is responsible
for maintaining the strict local causality in the world we live in. These considerations lead us to the following remarkable
theorem:

\smallskip

{\parindent 0pt
\underbar{\bf Theorema Egregium:}}

{\parindent 0pt
{\bf Every quantum mechanical correlation can be understood as a classical,
local-realistic correlation among a set of points of a parallelized 7-sphere, represented by maps of the form}}
\begin{equation}
\pm\,1\,=\,{\mathscr A}
({\bf n},\,\lambda): {\rm I\!R}^3\!\times\Lambda\longrightarrow S^7 \hookrightarrow{\rm I\!R}^8.\label{theoequa}
\end{equation}
The corresponding physical picture is the same as in Figure \ref{fig-33}, but with ${S^3}$ replaced by ${S^7}$,
the 4D ball replaced by the 8D ball, and the number of statistically significant
measurement events, ${\mathscr A}$, ${\mathscr B}$, ${\mathscr C}$, ${\mathscr D}$, {\it etc.}, generalized to an arbitrary
number. It is important to note, however, that despite appearances neither ${S^3}$ nor ${S^7}$ is a round sphere. The Riemann
curvature of both ${S^3}$ and ${S^7}$ is zero, because they are both parallelized spheres \cite{What-1}.

${S^7}$, however, has a much richer topological structure than ${S^3}$.
As noted above, it happens to be homeomorphic to the space of unit octonions, which are
well known to form the most general division algebra possible. In the language of fiber bundles one can thus view a 7-sphere as
a 4-sphere worth of 3-spheres. Each of its fiber is then a 3-sphere, and each one of these 3-spheres is a 2-sphere
worth of circles. Thus the four parallelizable spheres---${S^0}$, ${S^1}$, ${S^3}$, and ${S^7}$---can all be viewed as nested
within a 7-sphere. The EPR-Bohm correlations can then be understood as correlations among the equatorial points of one of the
fibers of this 7-sphere, as we saw above. Alternatively, a 7-sphere can be thought of as a 6-sphere worth of circles.
Thus the above theorem can be framed entirely in terms of circles, each one of which described by a classical, octonionic spinor
with a well-defined sense of rotation (i.e., whether it describes a clockwise rotation about a point within the 7-sphere or a
counterclockwise rotation). This sense of rotation in turn defines a definite handedness (or orientation) about every point
of the 7-sphere. If we designate this handedness by a random number ${\lambda=\pm\,1}$,
then local measurement results for any
physical scenario can be represented by raw scores of the form
\begin{equation}
S^7\ni {\mathscr A}({\bf a},\,{\boldsymbol\mu})\,=\,(-\,J\cdot{{\bf N(a)}}\,)\,(+\,{\boldsymbol\mu}\cdot{{\bf N(a)}}\,)\,=\,
\begin{cases}
+\,1\;\;\;\;{\rm if} \!\!&{\boldsymbol\mu}\,=\,+\,J \\
-\,1\;\;\;\;{\rm if} \!\!&{\boldsymbol\mu}\,=\,-\,J,
\end{cases} \label{raw7-0}
\end{equation}
where ${{\bf a}\in{\rm I\!R}^3}$ and ${{\bf N(a)}\in{\rm I\!R}^7}$ are unit vectors, and ${{\boldsymbol\mu}=\lambda\,J}$ is
the hidden variable analogous to ${{\boldsymbol\mu}=\lambda\,I}$ with ${I={\bf e}_x}{{\bf e}_y}{{\bf e}_z}$ replaced by
\begin{equation}
J=\,{{\bf e}_1}{{\bf e}_2}{{\bf e}_4}\,+\,{{\bf e}_2}{{\bf e}_3}{{\bf e}_5}\,+\,{{\bf e}_3}{{\bf e}_4}{{\bf e}_6}\,+
\,{{\bf e}_4}{{\bf e}_5}{{\bf e}_7}\,+\,{{\bf e}_5}{{\bf e}_6}{{\bf e}_1}\,+\,{{\bf e}_6}{{\bf e}_7}{{\bf e}_2}\,+
\,{{\bf e}_7}{{\bf e}_1}{{\bf e}_3}.
\end{equation}
The standard scores corresponding to these raw scores are then given by ${{\boldsymbol\mu}\cdot{\bf N(a)}}$, which
geometrically represent the equatorial points of a parallelized 7-sphere, just as ${{\boldsymbol\mu}\cdot{\bf a}}$
represented the equatorial points of a parallelized 3-sphere. In Chapters 6 and 7 we shall see explicit examples
of the functions ${\bf N(a)\in{\rm I\!R}^7}$ corresponding to measurement directions ${{\bf a}\in{\rm I\!R}^3}$.
The correlations between the raw scores
are then calculated as covariance between the standard scores ${{\boldsymbol\mu}\cdot{\bf N(a)}}$. Note also
that, just as in the EPR case, both the raw scores (\ref{raw7-0}) and the standard
scores ${{\boldsymbol\mu}\cdot{\bf N(a)}}$ are manifestly non-contextual.

Equipped with these variables, we are now in a position to prove the above theorem \cite{What-1}. To this end, recall that
no matter which model of physics we are concerned with---the quantum mechanical model, the
hidden variable model, or any other---for theoretical purposes all we need to understand are the expectation values of the
observables measured in various states of the physical systems \cite{von-1}. Accordingly,
consider an arbitrary quantum state ${|\Psi\rangle\in{\cal H}}$, where
${\cal H}$ is a Hilbert space of arbitrary dimensions, which may or may not be finite. We impose no restrictions
on either ${|\Psi\rangle}$ or ${\cal H}$, apart from their usual quantum mechanical meanings. In particular, the state
${|\Psi\rangle}$ can be as entangled as one may like, and the space ${\cal H}$ can be as large or small as one may like.
Next consider a self-adjoint operator ${{\cal\widehat O}({\bf a},\,{\bf b},\,{\bf c},\,{\bf d},\,\dots\,)}$ on this Hilbert
space, parameterized by an arbitrary number of local contexts ${{\bf a},\,{\bf b},\,{\bf c},\,{\bf d},}$ {\it etc.}
The quantum mechanical expectation value of this observable in the state ${|\Psi\rangle}$ is then given by:
\begin{equation}
{\cal E}_{{\!}_{Q.M.}}({\bf a},\,{\bf b},\,{\bf c},\,{\bf d},\,\dots\,)\,
=\,\langle\Psi|\;{\cal\widehat O}({\bf a},\,{\bf b},\,{\bf c},\,{\bf d},\,\dots\,)\,|\Psi\rangle\,.
\end{equation}
More generally, if the system happens to be in a mixed state, then
\begin{equation}
{\cal E}_{{\!}_{Q.M.}}({\bf a},\,{\bf b},\,{\bf c},\,{\bf d},\,\dots\,)\,
=\,\text{Tr}\left\{{W}\,{\cal\widehat O}({\bf a},\,{\bf b},\,{\bf c},\,{\bf d},\,\dots\,)\right\},\label{qmexpta}
\end{equation}
where ${W}$ is a statistical operator of unit trace representing the state.

Our goal now is to show that this expectation value can always be reproduced as a local-realistic expectation value of a set of
binary points of a parallelized 7-sphere. To this end, let
\begin{equation}
\pm\,1\,=\,{\mathscr A}_{\bf a}(\lambda): {\rm I\!R}^3\!\times\Lambda\longrightarrow {S^7}, \,\;\;
\pm\,1\,=\,{\mathscr B}_{\bf b}(\lambda): {\rm I\!R}^3\!\times\Lambda\longrightarrow {S^7},\;\text{\it etc.}
\end{equation}
be the raw scores of the form (\ref{raw7-0}). Using prescriptions analogous
to (\ref{var-a}) the corresponding standard scores then work out to be
\begin{equation}
{\boldsymbol\mu}\cdot{\bf N(a)}: {\rm I\!R}^3\!\times\Lambda\longrightarrow {S^6}, \;\;\;
{\boldsymbol\mu}\cdot{\bf N(b)}: {\rm I\!R}^3\!\times\Lambda\longrightarrow {S^6},\;\text{\it etc.} \label{stansco}
\end{equation}
Here ${\bf N(a)}$, ${\bf N(b)}$,${\;}${\it etc.}${\;}$may not necessarily be the same function for all
${{\bf n}\in{\rm I\!R}^3}$. They may be different functions for different directions.

This idealized prescription of raw scores and standard scores can, and should, be further generalized. So far we have
presumed that randomness in these scores originates entirely from the initial state ${\lambda}$ representing the
orientation of the 7-sphere. In other words, we have presumed that the local interactions of the measuring devices
${(-\,J\cdot{{\bf N(a)}}\,)}$ with the physical variables ${{\boldsymbol\mu}\cdot{{\bf N(a)}}}$ do not introduce
additional randomness in the scores ${{\mathscr A}_{\bf a}(\lambda)}$. Any realistic interaction between
${(-\,J\cdot{{\bf N(a)}}\,)}$ and ${{\boldsymbol\mu}\cdot{{\bf N(a)}}}$, however, {\it would} introduce such a
randomness of purely local origin. We can model it by an additional\break random variable ${\nu=\pm\,1}$
with probability ${0\leq p(\,\nu\,|\,{\bf a},\,\lambda\,)\leq 1\,}$, so that the
bivectors ${(-\,J\cdot{{\bf N(a)}}\,)}$ representing the measuring devices may now also take the random form
${(-\,\nu\,J\cdot{{\bf N(a)}}\,)}$.
The average of the corresponding raw scores ${{\mathscr A}_{\bf a}(\lambda)=\pm\,1}$ would then satisfy
\begin{equation}
-\,1\,\leq\,{\overline{{\mathscr A}_{\bf a}(\lambda)}}\,\leq\,+\,1\,,
\end{equation}
${}$
\vspace{-20pt}
\begin{align}
\text{with}\;\;\;\;\;{\overline{{\mathscr A}_{\bf a}(\lambda)}}\,&=\,\sum_{\nu}\, p(\,\nu\,|\,{\bf a},\,\lambda\,)
\,{\mathscr A}_{\bf a}(\nu,\,\lambda)\, \notag \\
&=\,\left[\sum_{\nu}\, p(\,\nu\,|\,{\bf a},\,\lambda\,)\,(-\,\nu\,J\cdot{{\bf N(a)}}\,)\right]
(+\,{\boldsymbol\mu}\cdot{{\bf N(a)}}\,)\;\;\;\;\;\text{\;\;\;\;} \notag \\
&=\,{\overline{(-\,J\cdot{{\bf N(a)}}\,)}}\;(+\,{\boldsymbol\mu}\cdot{{\bf N(a)}}\,)\,.
\end{align}
Not surprisingly, this does not affect the corresponding standard scores (\ref{stansco}) worked out earlier. But if,
in addition, we assume that the common cause ${\lambda=\pm\,1}$ itself is also distributed non-uniformly between its values
${+\,1}$ or ${-\,1}$, then the standard scores modify to
\begin{equation}
{\boldsymbol\mu}\cdot{\bf N(a)}\longrightarrow  \kappa(\lambda)\,{\boldsymbol\mu}\cdot{\bf N(a)}, \;\;\;
{\boldsymbol\mu}\cdot{\bf N(b)}\longrightarrow  \kappa(\lambda)\,{\boldsymbol\mu}\cdot{\bf N(b)}\;\text{\it etc.},
\end{equation}
${}$
\vspace{-13pt}
\begin{equation}
\!\!\!\!\!\!\!\!\!\!\!\!\!\!\!\!\!\!
\!\!\!\!\!\!\!\!\!\!\!\!\!\!\!\!\!\!\!\!\!\!\!\!\!\!\!\!\!\!\!\!\!\!\!\!
\text{where}\;\;\;\;\;\;\;\;\;\;\;\;\;\;\;\;\;\;\;\;\;\;\;\;\;\;
\kappa(\lambda)\,=\,\frac{1-\lambda\,{\overline{\lambda}}}{\sqrt{1-\left(\,{\overline{\lambda}}\,\right)^2\,}\;}\,,\;\;\;\;
\end{equation}
with ${\overline{\lambda}}$ being the average over the probability distribution of ${\lambda}$.
The correlation among the raw scores ${{\mathscr A}_{\bf a}(\lambda)=\pm\,1}$, ${{\mathscr B}_{\bf b}(\lambda)=\pm\,1}$,
${{\mathscr C}_{\bf c}(\lambda)=\pm\,1}$, {\it etc.}${\;}$can now be easily calculated as covariance among the standard scores
${{A}_{\bf a}(\lambda)={\boldsymbol\mu}\cdot{\bf N(a)}}$, ${{B}_{\bf b}(\lambda)={\boldsymbol\mu}\cdot{\bf N(b)}}$, {\it etc.} as
\begin{equation}
{\cal E}_{{\!}_{L.R.}}({\bf a},\,{\bf b},\,{\bf c},\,{\bf d},\,\dots\,)\,=\int_{\Lambda}
{A}_{\bf a}(\lambda)\,{B}_{\bf b}(\lambda)\,{C}_{\bf c}(\lambda)\,\dots\,\;\rho(\lambda)\;d\lambda\,,\label{locnest}
\end{equation}
where the overall probability distribution ${\rho(\lambda)=\kappa^m(\lambda)}$ is allowed to be both a non-uniform and
continuous function of ${\lambda}$, with ${m}$ being the total number of local contexts
${{\bf a},\,{\bf b},\,{\bf c},\,{\bf d},\,\dots}$ in the experiment.

To evaluate these correlations, note that the standard scores ${{A}_{\bf a}(\lambda)={\boldsymbol\mu}\cdot{\bf N(a)}}$,
${{B}_{\bf b}(\lambda)={\boldsymbol\mu}\cdot{\bf N(b)}}$, ${{C}_{\bf c}(\lambda)={\boldsymbol\mu}\cdot{\bf N(c)}}$,
{\it etc.}${\;}$are in fact bivectors representing the equatorial points of the 7-sphere, which remains as closed under
multiplication as the 3-sphere. As a\break result, the product of the standard scores can be written as
\begin{equation}
{A}_{\bf a}(\lambda)\,{B}_{\bf b}(\lambda)\,{C}_{\bf c}(\lambda)\dots=
f({\bf a},\,{\bf b},\,{\bf c},\,\dots\,)+
{P}_{\bf \widehat{N}}(\lambda)\,\;g\,({\bf a},\,{\bf b},\,{\bf c},\dots)\,,\label{reoct}
\end{equation}
where the RHS is an octonionic spinor representing a non-equatorial point of the 7-sphere, the vector
${{\bf \widehat{N}}({\bf a},\,{\bf b},\,{\bf c},\,{\bf d},\,\dots\,)\in{\rm I\!R}^7}$ is a function of all
3D vectors, ${{P}_{\bf \widehat{N}}(\lambda)\equiv
{\boldsymbol\mu}\cdot{\bf \widehat{N}({\bf a},\,{\bf b},\,{\bf c},\,{\bf d},\dots)}}$ is a unit bivector
representing an equatorial point of ${S^7}$ that is different from the ones represented by the bivectors
${{A}_{\bf a}(\lambda)}$, ${{B}_{\bf b}(\lambda)}$, ${{C}_{\bf c}(\lambda)}$, {\it etc.}, and the scalar
functions ${f({\bf a},\,{\bf b},\,{\bf c},\,\dots\,)}$ and
${g({\bf a},\,{\bf b},\,{\bf c},\,\dots\,)}$ satisfy ${f^2\!+g^2=1}$ with
${f}$ identified as the quantum mechanical expectation value (\ref{qmexpta}):
\begin{equation}
f({\bf a},\,{\bf b},\,{\bf c},\,{\bf d},\,\dots\,)\,=\,
\text{Tr}\left\{{W}\,{\cal\widehat O}({\bf a},\,{\bf b},\,{\bf c},\,{\bf d},\,\dots\,)\right\}.
\end{equation}
Conversely, any arbitrary point of the 7-sphere (or joint beable)
\begin{equation}
(\,{A}_{\bf a}\,{B}_{\bf b}\,{C}_{\bf c}\,{D}_{\bf d}\dots)(\lambda)
\,=\,f({\bf a},\,{\bf b},\,{\bf c},\,\dots\,)+
{P}_{\bf \widehat{N}}(\lambda)\,\;g\,({\bf a},\,{\bf b},\,{\bf c},\dots)
\end{equation}
corresponding to the quantum mechanical operator ${{\cal\widehat O}({\bf a},\,{\bf b},\,{\bf c},\,{\bf d},\dots)}$
can always be factorized into any number of local parts as
\begin{equation}
S^7\,\ni\,(\,{A}_{\bf a}\,{B}_{\bf b}\,{C}_{\bf c}\,{D}_{\bf d}\,\dots\,)(\lambda)\,=\,                          
{A}_{\bf a}(\lambda)\,{B}_{\bf b}(\lambda)\,{C}_{\bf c}(\lambda)\,                                           
{D}_{\bf d}(\lambda)\,\dots\,,
\end{equation}
since, as we have already noted, the 7-sphere remains closed under multiplication of any number of its points.
Using the identity (\ref{reoct}), the local realistic expectation value (\ref{locnest}) can now be rewritten as
\begin{align}
{\cal E}_{{\!}_{L.R.}}({\bf a},\,{\bf b},\,{\bf c},\,{\bf d},\,&\dots\,)\,=
\,f({\bf a},\,{\bf b},\,{\bf c},\,{\bf d},\,\dots\,)\int_{\Lambda}\rho(\lambda)\;d\lambda \notag \\
&\;\,+\,g\,({\bf a},\,{\bf b},\,{\bf c},\,{\bf d},\,\dots\,)\int_{\Lambda}\,{P}_{\bf \widehat{N}({\bf n})}(\lambda)
\,\;\rho(\lambda)\;d\lambda\,.\label{moldpop}
\end{align}
Note, however, that the vector ${{\bf \widehat{N}}({\bf n})\in{\rm I\!R}^7}$ appearing in the second term here corresponds
to a 3D direction ${{\bf n}\in{\rm I\!R}^3}$ that is necessarily exclusive to all the other measurement directions
${{\bf a},\,{\bf b},\,{\bf c},\,{\bf d},\dots}$ As a result, the bivector
${{P}_{\bf \widehat{N}({\bf n})}(\lambda)={\boldsymbol\mu}\cdot{\bf \widehat{N}({\bf n})}}$ necessarily
corresponds to a null measurement result, reducing the second integral in (\ref{moldpop}) to zero \cite{What-1}\cite{illusion-1}.
If we next assume that the probability distribution, although not necessarily uniform, remains normalized to unity,
\begin{equation}
\int_{\Lambda}\rho(\lambda)\;d\lambda\,=\,1\,,
\end{equation}
then the above expectation value reduces to
\begin{equation}
{\cal E}_{{\!}_{L.R.}}({\bf a},\,{\bf b},\,{\bf c},\,{\bf d},\,\dots\,)\,=\,
f({\bf a},\,{\bf b},\,{\bf c},\,{\bf d},\,\dots\,)\,.
\end{equation}
This finally proves our main theorem: Every quantum mechanical correlation can be
understood as a classical, local-realistic correlation among a set of points of a parallelized 7-sphere. Q.E.D.}

{\baselineskip 11.2pt

\section{\large The Raison D'\^etre of Quantum Correlations}

The above result demonstrates that the discipline of parallelization in the manifold of all possible measurement results is
responsible for the existence and strength of {\it all}
${\,}$quantum correlations. More precisely, it identifies quantum correlations
as evidence that the physical space we live in respects the symmetries and topologies of a parallelized 7-sphere. As we shall
see in greater detail in Chapter 7, there are profound mathematical and conceptual reasons why the topology of the 7-sphere
plays such a significant role in the manifestation of quantum correlations. Essentially it is because 7-sphere happens to be
homeomorphic to the most general possible division algebra. And it is the property of division that turns out to be responsible
for maintaining strict local causality in the world we live in.

To understand this chain of reasoning better, recall that, just as a parallelized 3-sphere is a 2-sphere worth of
1-spheres but with a twist in
the manifold ${S^3\;(\not=S^2\times S^1)}$, a parallelized 7-sphere is a 4-sphere worth of 3-spheres but with a twist in the
manifold ${S^7\;(\not=S^4\times S^3)}$. More precisely, just as ${S^3}$ is a nontrivial fiber bundle over ${S^2}$ with
Clifford parallels ${S^1}$ as its linked fibers, ${S^7}$ is also a nontrivial fiber bundle, but over
${S^4}$, and with entire 3-dimensional spheres ${S^3}$ as its linked fibers. Now it is the twist in the
bundle ${S^3}$ that forces one to forgo the commutativity of complex numbers (corresponding to the circles ${S^1}$)
in favor of the non-commutativity of quaternions.
In other words, a 3-sphere is not parallelizable by the commuting complex numbers but only by the
non-commuting quaternions. In a similar vein, the twist in the bundle ${S^7\not=S^4\times S^3}$ forces one to forgo
the associativity of quaternions (corresponding to the fibers ${S^3}$) in favor of the non-associativity of
octonions. In other words, a 7-sphere is not parallelizable by the associative quaternions but only by the
non-associative octonions. And the reason why it can be parallelized at all is because
its tangent bundle happens to be trivial:
\begin{equation}
{\rm T}S^7\,=\!\bigcup_{\,p\,\in\, S^7}\{p\}\times T_pS^7\,\equiv\,S^7\times{\rm I\!R}^7.
\end{equation}

Once parallelized by a set of unit octonions, both the 7-sphere and each of its 3-spherical fibers remain closed
under multiplication. This, in turn, means that the factorizability or locality condition of Bell is automatically
satisfied within a parallelized 7-sphere. The lack of associativity of octonions, however, entails that, unlike
the unit 3-sphere (which is homeomorphic to the Lie group SU(2)), a 7-sphere is not a group manifold, but forms only
a quasi-group. As a result, the torsion within the 7-sphere continuously varies from one point to another of the
manifold \cite{What-1}. It is this variability of the parallelizing\break torsion within ${S^7}$ that is ultimately responsible
for the diversity and non-linearity of the quantum correlations we observe in nature:
\begin{center}
Parallelizing Torsion ${\,{\cal T}_{\,\alpha\,\beta}^{\,\gamma}\not=0}$ ${\;\;\;\,\Longleftrightarrow\;\;\;}$
Quantum Correlations.
\end{center}
The upper bound on all possible quantum correlations is thus set by the maximum of possible torsion within the 7-sphere:
\begin{center}
Maximum of Torsion ${{\cal T}_{\,\alpha\,\beta}^{\,\gamma}\not=0}$ ${\;\;\;\;\Longrightarrow\,\;\;}$ The Upper Bound
${\,2\sqrt{2}}$.
\end{center}
These last two results will be proved rigorously in Chapter 7.

\section{\large Local Causality and the Division Algebras}

In the last few sections we saw the crucial role played by the 3- and 7-dimensional spheres in understanding the existence
of quantum correlations. What is so special about 3 and 7 dimensions? Why is the vector cross product definable only in 3 and 7
dimensions and no other? Why are ${\mathbb R}$, ${\mathbb C}$, ${\mathbb H}$, and ${\mathbb O}$ the only possible normed division
algebras? Why are only the 3- and 7-dimensional spheres nontrivially parallelizable out of infinitely many possible spheres? Why
is it possible to derive all quantum mechanical correlations as local-realistic correlations among the points of only
the 7-sphere?

The answers to all of these questions are intimately connected to the notion of factorizability introduced by
Bell within the context of his theorem. Mathematicians have long been asking: When is a product of two squares itself a square:
${x^2\,y^2 = z^2\,}$? If the number ${z}$ is factorizable, then it can be written as a product of two other numbers, ${z = x\,y}$,
and then the above equality is seen to hold for the numbers ${x}$, ${y}$, and ${z}$. For ordinary numbers this is easy to check.
The number 8 can be factorized into a product of 2 and 4, and we then have ${64 = 8^2 = (2 \times 4)^2 = 2^2\times 4^2 = 64}$.
But what about sums of squares? A more profound equality holds, in fact, for a sum of two squares times a sum of
two squares as a third sum of two squares:
\begin{equation}
(x_1^2 + x_2^2)\,(y_1^2 + y_2^2) \,=\, (x^{}_1y^{}_1 - x^{}_2y^{}_2)^2 + (x^{}_1y^{}_2 + x^{}_2y^{}_1)^2\,=\,z_1^2 + z_2^2.
\end{equation}
There is also an identity like this one for the sums of four squares. It was first discovered by Euler, and then rediscovered and
popularized by Hamilton in the 19th
century through his work on quaternions. It is also known that Graves and Cayley independently discovered a similar identity for
the sums of eight squares. This naturally leads to the question of whether the product of two sums of squares of ${n}$ different
numbers can be a sum of ${n}$ different squares? In other words, does the following equality hold in general for any ${n}$?
\begin{equation}
(x_1^2 + x_2^2 + \dots + x_n^2)\,(y_1^2 + y_2^2 + \dots + y_n^2) \,=\, z_1^2 + z_2^2 + \dots + z_n^2.\label{8.39}
\end{equation}
It turns out that this equality holds only for ${n}$ = 1, 2, 4, and 8. This was proved by Hurwitz in 1898 \cite{Hurwitz-rrr}.
It reveals a deep and surprising fact about the world we live in. Much of what we see around us, from elementary particles to
distant galaxies, is an inevitable consequence of this simple mathematical fact. The world is the way it is because the above
equality holds only for ${n}$ = 1, 2, 4, and 8. For example, the above identity is equivalent to the existence of a division
algebra of dimension ${n}$ over the field ${\mathbb R}$ of real numbers.
Indeed, if we define vectors ${{\bf x}=(x_1,\dots, x_n)}$, ${{\bf y}=(y_1,\dots, y_n)}$, and
${{\bf z}={\bf x}*{\bf y}}$
in ${{\mathbb R}^n}$ such that ${z_i}$'s are functions of ${x_j}$'s and ${y_k}$'s determined by (\ref{8.39}), then
\begin{equation}
||{\bf x}||\,||{\bf y}|| \,=\, ||{\bf x}*{\bf y}||\,.
\end{equation}
Thus the division algebras ${\mathbb R}$ (real), ${\mathbb C}$ (complex), ${\mathbb H}$ (quaternion), and ${\mathbb O}$
(octonion) we use in much of our science are intimately related to the dimensions ${n}$ = 1, 2, 4, and 8. 
Moreover, from the equation of a unit sphere,
\begin{equation}
x_1^2 + x_2^2 + \dots + x_n^2 \,=\,1\,,
\end{equation}
it is easy to see that the four parallelizable
spheres ${S^0}$, ${S^1}$, ${S^3}$, and ${S^7}$ correspond to ${n}$ = 1, 2, 4, and 8, which are the dimensions of
the respective embedding spaces of these four spheres. What is not so easy to see, however, is the fact that
there is a deep connection between Hurwitz's theorem and the quantum correlations \cite{What-1}.
As we saw in the previous sections, all quantum correlations are inevitable consequences of the parallelizability of the
7-sphere, which in turn is a consequence of Hurwitz's theorem. So the innocent looking algebraic equality (\ref{8.39})
has far reaching consequences, not only for the entire edifice of mathematics, but also for that of quantum physics.

\section{\large Concluding Remarks}

The results obtained in the preceding sections (and in the chapters that are to follow \cite{What-1})
go far deeper and well beyond the boundaries of Bell's theorem. In any physical experiment what is observed
are ``clicks'' of the event detectors corresponding to yes/no answers to our questions.
As in the EPR experiment \cite{Weihs-1}\cite{disproof-1}, when we compare the answers recorded by various
observers in quantum experiments conducted at mutually remote locations,
we find that their answers are correlated in a mathematically and statistically very disciplined manner.
The natural question then clearly is: why are these answers correlated in such a disciplined manner
when there appears to be no predetermined common cause dictating the correlations. Bell and his followers claimed
that the observed correlations are the evidence of a radical non-locality in nature. The stronger adherents of this view
often claim that ``nature is non-local.'' Here we have rejected this view. Instead, we have demonstrated how a perfectly
natural local explanation of the observed correlations is possible. In our view these correlations are the evidence,
not of non-locality, but the fact that the physical space we live in respects the symmetries and topologies of
a parallelized 7-sphere \cite{What-1}. The 3-dimensional space we normally presume as our reality is then
simply one of the many fibers of this\break 7-sphere. Our observations are still confined to various
3-dimensional subspaces of this 7-sphere, but the correlations among the results of our experiments are
revealing that the observed measurement events are actually occurring within a 7-dimensional manifold. The radical
non-locality of Bell is thus traded off for the extra dimensions going beyond our immediate experiences in the
macroscopic world.

The methodology that has led us to this conclusion is similar to that used by Einstein to arrive at his local field
theory of gravity. Just as the demand of locality in the face of Newton's non-local theory of gravity led Einstein to
general relativity, the demand of locality in the face of quantum correlations has led us to a parallelized 7-sphere.

\vspace{0.5cm}

{\parindent 0pt

{\large\bf Acknowledgments}

\smallskip

I wish to thank David Coutts, Fred Diether, Azhar Iqbal, Edwin Eugene Klingman, Rick Lockyer,
Ray B. Munroe, and Tom H. Ray for discussions.
I also wish to thank the Foundational Questions Institute (FQXi) for supporting this work through a Mini-Grant.}

\vspace{0.5cm}

{\parindent 0pt

{\large\bf Appendix 1: A 2-D Analogue of EPRB Correlation}

\vspace{0.5cm}

Suppose Alice and Bob are two-dimensional creatures living in a two-dimensional, one-sided world resembling a M\"obius strip,
entirely oblivious to the third dimension we take for granted (cf. Fig. \ref{fig-not-1}). Suppose further that they discover
certain correlation between the results of their observations that appears to be much stronger than any previously observed
correlation, and its strength appears to be explainable only in non-local terms. Alice and Bob may, however, strive to discover
a hypothesis that could explain the correlation in purely local terms. They may hypothesize, for example, that they are in fact
living in a M\"obius world, embedded in a higher-dimensional space ${{\rm I\!R}^3}$.
The purpose of this appendix is to illustrate how such a
hypothesis would explain their observed correlation in purely local terms, and relate it to the hypothesis we have advanced above
to explain the correlations we observe in our three-dimensional world.}

To this end, let Alice and Bob choose directions ${\bf a}$ and ${\bf b}$ to perform two independent set of experiments,
conducted at remote locations from each other. Within their two-dimensional
world the vectors ${\bf a}$ and ${\bf b}$
could only have two coordinate components, but Alice and Bob hypothesize that perhaps the vectors also have third
components, pointing ``outside'' of their own one-sided world. Let
the twisting angle between these external components be denoted by ${\eta_{\scriptscriptstyle{\bf ab}}}$,
as shown in Fig.${\;}$\ref{fig-not-2} (b) below.
Now the experiments Alice and Bob have been performing are exceedingly simple.
It so happens that within their two-dimensional world wherever they set up their
posts and choose directions for their measurements, they start receiving a stream of L-shaped patterns. Upon receiving each such
pattern they record whether it has a left-handed L-shape or a right-handed L-shape. They determine this by aligning the longer
arm of each pattern along their chosen measurement direction, with the shorter arm hanging in the opposite direction, as shown
in Fig.${\;}$\ref{fig-not-1}. It is then easy for them to see whether the pattern has a left-handed L-shape
or a right-handed L-shape. If it turns out to have a left-handed
L-shape, Alice and Bob record the number ${-1}$ in their logbooks, and if it turns
out to have a right-handed L-shape, they record the number ${+1}$ in their logbooks. What they always find in any such experiment
involving a large number of patterns is that the sum total of all the numbers they end up recording independently of each
other always add
up to zero. In other words, the L-shaped patterns they both
ceaselessly receive are always evenly distributed between the left-handed
patterns and the right-handed patterns. But when
they get together at the end of the day and compare the entries of their logbooks,
they find that their observations are strongly correlated. They find, in fact, that the correlation among their observations
can be expressed in terms of the angle ${\eta_{\scriptscriptstyle{\bf ab}}}$ (defined above) as
\begin{equation}
{\cal E}({\bf a},\,{\bf b})\,=\,-\cos\eta_{\scriptscriptstyle{\bf ab}}\,.\label{exeta}
\end{equation}

\begin{figure}
\hrule
\vspace{-0.85cm}
\scalebox{0.9}{
\begin{pspicture}(-2.6,-0.5)(2.0,5.2)

\epsfig{figure=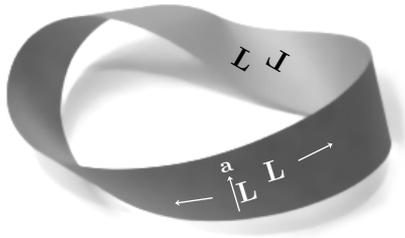,height=102pt,width=169.8pt}

\rput{-30}(-2.5,2.74){\large{\bf L}}

\rput{-30}(-2.0,2.74){\reflectbox{\large{\bf L}}}

\rput{17}(-2.05,1.11){\color{white}{\large{\bf L}}}

\rput{17}(-2.45,0.80){\color{white}{\large{\bf L}}}

\psline[linewidth=0.2mm,linecolor=white,arrowinset=0.3,arrowsize=1pt 3,arrowlength=0.7]{->}(-2.557,0.50)(-2.71,1.02)

\psline[linewidth=0.2mm,linecolor=white,arrowinset=0.3,arrowsize=1pt 3,arrowlength=0.7]{->}(-1.7,1.25)(-1.2,1.5)

\psline[linewidth=0.2mm,linecolor=white,arrowinset=0.3,arrowsize=1pt 3,arrowlength=0.7]{->}(-2.95,0.7)(-3.5,0.62)

\rput{17}(-2.75,1.16){\color{white}{\small ${\bf a}$}}

\end{pspicture}}
\vspace{-0.13cm}
\hrule
\caption{In the two-dimensional one-sided world of Alice and Bob two congruent shapes may become
incongruent relative to each other.\break}
%\vspace{-0.25cm}
\label{fig-not-1}
\hrule
\end{figure}
\begin{figure}
\hrule
\vspace{-0.25cm}
\scalebox{0.85}{
\begin{pspicture}(-3.4,-0.5)(1.0,5.2)

\put(-0.3,3.92){\huge${\infty}$}

\psline[linewidth=0.2mm,linecolor=white,arrowinset=0.3,arrowsize=1pt 3,arrowlength=0.7]{->}(-8.557,0.50)(-8.71,1.02)

\rput{17}(-8.75,1.16){\color{white}{\small ${\bf a}$}}

\psarc[linewidth=0.4mm,arrowinset=0.3,arrowsize=2pt 4,arrowlength=0.8]{>-<}(0.1,2.1){2.0}{99.12}{82.98}

\put(0.7,2.28){\large${\beta_{\scriptscriptstyle{\bf ab}}}$}

\psarc[showpoints=true,linewidth=0.2mm,linestyle=none]{*-*}(0.11,2.11){1.995}{0}{40}

\psarc[arrowinset=0.3,arrowsize=1pt 3,arrowlength=1]{->}(0.1,2.1){1.3}{0}{40}

\put(2.25,2.0){\Large ${\mathscr A}$}

\put(1.7,3.55){\Large ${\mathscr B}$}

\put(-0.1,0.9){\large ${{\rm I\!R}^3}$}

\put(-0.1,-0.6){\large (a)}

\psline[linewidth=0.5mm,arrowinset=0.3,arrowsize=1pt 3,arrowlength=1]{<-}(5.5,3.7)(5.5,0.6)

\psline[linewidth=0.5mm,arrowinset=0.3,arrowsize=1pt 3,arrowlength=1]{<-}(6.6,3.2)(4.4,1.0)

\psline[linewidth=0.3mm,linestyle=dashed,arrowinset=0.3,arrowsize=1pt 3,arrowlength=1]{->}(5.5,2.1)(6.75,2.1)

\psline[linewidth=0.3mm,linestyle=dashed,arrowinset=0.3,arrowsize=1pt 3,arrowlength=1]{->}(5.5,2.1)(6.4,1.2)

\put(6.9,2.02){\large ${\rm z}$}

\rput{-50}(6.57,1.06){\large ${\rm z}$}

\put(5.37,3.93){\large ${\bf a}$}

\put(6.75,3.2){\large${\bf b}$}

\psarc[arrowinset=0.3,arrowsize=1pt 3,arrowlength=1]{<-}(5.5,2.3){0.6}{31.2}{90}

\put(5.7,3.1){\large${\eta_{\scriptscriptstyle{\bf ab}}}$}

\put(5.25,-0.6){\large (b)}

\end{pspicture}}
\vspace{0.7cm}
\hrule
\caption{(a) The aerial view of the M\"obius world of Alice and Bob. The distance between their observation posts is
given by the angle ${0\leq\beta_{\scriptscriptstyle{\bf ab}}\leq 2\pi}$. (b) The cross-sectional view of the M\"obius world. The
torsional twist in the strip is characterized by
the angle ${0\leq\eta_{\scriptscriptstyle{\bf ab}}\leq\pi}$.\break}
\label{fig-not-2}
\hrule
\end{figure}
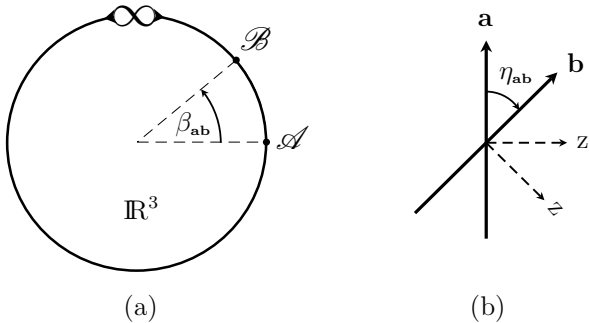

To explain this correlation in purely local terms Alice and Bob hypothesize that the three-dimensional space external
to their own is occupied by a mischievous gremlin, who is hurling complementary L-shaped patterns towards them in a
steady stream. What is more, this gremlin has a habit of making a random but evenly distributed choice between hurling
a pattern towards his right or his left, with a complementary pattern hurled in the opposite direction. Each choice of
the gremlin thus constitutes an evenly distributed random  hidden variable ${\lambda=\pm\,1}$, determining both the
initial state of the patterns as well as the measurement outcomes of Alice and Bob:
\begin{align}
\;\,{\mathscr A}({\bf a},\,{\lambda})\,&=\,
\begin{cases}
+\,1\;\;\;\;\;{\rm if} &\lambda\,=\,+\,1 \\
-\,1\;\;\;\;\;{\rm if} &\lambda\,=\,-\,1
\end{cases} \label{mob-1}
\end{align}
\begin{align}
\text{and}\;\;\;\;
{\mathscr B}({\bf b},\,{\lambda})\,&=\,
\begin{cases}
-\,1\;\;\;\;\;{\rm if} &\lambda\,=\,+\,1 \\
+\,1\;\;\;\;\;{\rm if} &\lambda\,=\,-\,1\,.\;\;\;\;\;\;
\end{cases} \label{mob-2}
\end{align}
These outcomes, in addition to being local and deterministic, are also statistically independent events. What is more, they are
determined manifestly non-contextually. Since the numbers ${\mathscr A}$ and ${\mathscr B}$ are read off simply by aligning
the received patterns along the chosen vectors ${\bf a}$ and ${\bf b}$ and noting whether the patterns are right- or left-handed,
the values or directions of these vectors are of no real significance for the measurements of Alice and Bob. Any local direction
${\bf a}$ chosen by Alice, for example, would give the same answer to the question whether a
received pattern is right-handed or left-handed. Moreover, the z-components of ${\bf a}$ and ${\bf b}$
(inaccessible to Alice and Bob) also have no significance as far as their local measurements are concerned. 

Now it may appear from the definitions (\ref{mob-1}) and (\ref{mob-2}) that the product ${{\mathscr A}{\mathscr B}}$ of the
results of Alice and Bob will always remain at the fixed value of ${-1}$, but that, in fact, is an illusion. The product,
in fact, will fluctuate inevitably between the values ${-1}$ and ${+1}$,
\begin{equation}
{\mathscr A}{\mathscr B} \,\in\, \{\,-1,\,+1\,\},
\end{equation}
not the least because ${\mathscr A}$ and ${\mathscr B}$ are statistically independent events.
To appreciate this, recall that in a one-sided world of a M\"obius strip two congruent left-handed figures may not
remain congruent for long (cf. Fig.${\;}$\ref{fig-not-1}).
If one of the two left-handed figures moves around the strip {\it relative} to the
stay-at-home figure it becomes right-handed, and hence incongruent with the stay-at-home figure. And the same is true for
two right-handed figures\footnote{This is analogous to the real world fact that bivectors are not isolated objects${\,}$
but represent relative rotations within the constraints of the physical space \cite{What-1}.}.
This is quite easy to check by making a model of a M\"obius strip from a strip of paper.
If one starts with two congruent L-shaped cardboard cutouts and moves one of them around the strip relative to the
other, then after a complete revolution the two cutouts become incongruent with one another. As a result the
value of the corresponding product ${{\mathscr A}{\mathscr B}}$ representing congruence or incongruence of the
two L-shaped figures would change from ${+1}$ at the start of the
trip to ${-1}$ at the end of the trip. 

The reason for this of course is that there is a torsional twist in the M\"obius strip similar to the one within the Clifford
parallels that constitute the 3-sphere \cite{What-1}. This twist is what is responsible for the strong correlation observed
by Alice and Bob. To understand this better, suppose the gremlin happens to hurl a right-handed pattern towards
Alice and a left-handed pattern towards Bob ({\it i.e.}, suppose ${\lambda=+1}$). What are the chances that Alice would
then record the number ${+1}$ in her logbook whereas Bob would record the number ${-1}$ in his logbook?
The answer to this question would depend, in fact, on where Alice and Bob are
situated within the M\"obius strip, which can be parameterized in terms of the external
angle ${\beta_{\scriptscriptstyle{\bf ab}}\,}$, as shown in Fig.${\;}$\ref{fig-not-2} (a). If the posts of Alice and Bob are
almost next to each other, then their patterns are unlikely to undergo relative handedness\break transformation, and then
Alice and Bob would indeed record ${+1}$ and ${-1}$, respectively, yielding the product ${{\mathscr A}{\mathscr B}=-1}$.
If, however, Bob's post is almost a full circle away from Alice's post, then both Alice and
Bob would record ${+1}$ with near certainty, because then Bob's pattern would have transformed into a right-handed pattern
relative to Alice's pattern with near certainty, yielding the product\break ${{\mathscr A}{\mathscr B}=+1}$.
For all intermediate angles the probability of the two patterns having undergone relative handedness transformation would be
equal to ${{\beta_{\scriptscriptstyle{\bf ab}}}/2\pi}$, and the probability of the same two patterns {\it not}\break having
undergone relative handedness transformation would be equal to ${(2\pi - {\beta_{\scriptscriptstyle{\bf ab}}})/2\pi}$.
Thus all four possible combinations of outcomes, ${+\,+}$, ${+\,-}$, ${-\,+}$, and ${-\,-}$, would be observed by Alice and Bob,
just as in equation (\ref{calcul}) discussed
above. The corresponding correlation among their measurement results would therefore be given by
\begin{align}
{\cal E}({\bf a},\,{\bf b})&\,=\,\lim_{\,n\,\gg\,1}\left[\frac{1}{n}\sum_{i\,=\,1}^{n}\,{\mathscr A}({\bf a},\,{\lambda}^i)\,
{\mathscr B}({\bf b},\,{\lambda}^i)\right] \notag \\
&\,=\,\frac{\beta_{\scriptscriptstyle{\bf ab}}}{2\pi}\times({\mathscr A}{\mathscr B}=+1)
\,+\, \frac{(2\pi-\beta_{\scriptscriptstyle{\bf ab}})}{2\pi}\times({\mathscr A}{\mathscr B}=-1) \notag \\
&\,=\,\frac{\beta_{\scriptscriptstyle{\bf ab}}\times(+1)\,+\,(2\pi-\beta_{\scriptscriptstyle{\bf ab}})\times(-1)}{2\pi} \notag \\
&\,=\,-1\,+\,\frac{1}{\pi}\;\beta_{\scriptscriptstyle{\bf ab}}\;\;\;\,\left(i.e.,\;\,{\rm linear\;\,within}
\;\,{\rm I\!R}^3\right)\!.\label{mobcorfi}
\end{align}
The validity of this result is straightforward to check by substituting ${\beta_{\scriptscriptstyle{\bf ab}}=0}$,
${\pi}$, and ${2\pi}$ to obtain ${{\cal E}({\bf a},\,{\bf b})=-1}$, ${0}$, and ${+1}$, respectively.

This correlation is expressed, however, in terms of the external angle ${\beta_{\scriptscriptstyle{\bf ab}}}$, which Alice
and Bob can measure in radians as a distance between their observation posts. As a final step towards explaining the
correlation (\ref{exeta}) they must now work out the distance ${\beta_{\scriptscriptstyle{\bf ab}}}$ in terms of the angle
${\eta_{\scriptscriptstyle{\bf ab}}}$, which is not a difficult task. Recalling the properties of a M\"obius strip
it is easy to see that these angles are in fact related as
\begin{equation}
\beta_{\scriptscriptstyle{\bf ab}}\,=\,\pi\left\{1-\cos\eta_{\scriptscriptstyle{\bf ab}}\right\}.\label{relaangle}
\end{equation}
This relation is the {\it defining} ${\,}$relation of the M\"obius world of Alice and Bob.
Substituting it into equation (\ref{mobcorfi}) they therefore obtain
\begin{align}
{\cal E}({\bf a},\,{\bf b})\,&=\,-1\,+\,\frac{1}{\pi}\;\beta_{\scriptscriptstyle{\bf ab}}\, \notag \\
&=\,-1\,+\,\frac{1}{\pi}\times\left[\,\pi\left\{1-\cos\eta_{\scriptscriptstyle{\bf ab}}\right\}\right] \notag \\
&=\,-\cos\eta_{\scriptscriptstyle{\bf ab}}\,.
\end{align}

Needless to say, as enlightening as it is, this fictitious analogue of the EPR-Bohm correlation cannot be taken too seriously.
It helps us to a certain extent in understanding the real world correlation within the 3-sphere, but there are analogies as well
as disanalogies between the two worlds. For example, although a torsional twist within the structure of both worlds is responsible
for the sinusoidal correlation, unlike the M\"obius strip a 3-sphere is an {\it orientable} ${\,}$manifold. Thus, it is not the
non-orientability, but the consistency of orientation within the 3-sphere that brings about the variations
${+\,+}$, ${+\,-}$, ${-\,+}$, and ${-\,-}$ in the observed results of the real Alice and Bob. Moreover, unlike in the
M\"obius world where relative handedness of two L-shapes depends on the distance between them, in the real world relative
handedness of the two bivectors appearing within the outcomes (\ref{17-noy-111})
and (\ref{18-noy-111}) reflects their intrinsic spinorial characteristics, independently of any distance between them. It is
constrained only by the identity
\begin{equation}
(\,{\boldsymbol\mu}\cdot{\bf a})(\,{\boldsymbol\mu}\cdot{\bf b})\,
=\,-\,{\bf a}\cdot{\bf b}\,-\,{\boldsymbol\mu}\cdot({\bf a}\times{\bf b}).
\end{equation}
This identity encapsulates the topology of a parallelized 3-sphere, with ${\boldsymbol\mu=\pm\,I}$ specifying its orientation.
It is clear from this identity that when ${{\bf b}=-\,{\bf a}}$ the handedness (about ${\bf a}$) of the bivectors
on its LHS differ from one another, whereas when ${{\bf b}=+\,{\bf a}}$ the handedness (about
${\bf a}$) is the same regardless of the distance between ${\bf a}$ and ${\bf b}$.
  
\vspace{0.5cm}

{\parindent 0pt

{\large\bf Appendix 2: The Meaning of a Geometric Product}

\vspace{0.5cm}

The concept of a geometric product was first introduced by Hermann Grassmann
to characterize what he called ``extensive magnitudes.''
Nowadays Hestenes refers to ``extensive magnitudes'' as ``directed numbers'' \cite{Clifford-1}.
To understand the meaning of the geometric product between two such directed numbers, let us first look at the ``inner product''
between two vectors, say ${\bf a}$ and ${{\bf b}\,}$:}
\begin{equation}
{\bf a}\cdot{\bf b}\,=\,\frac{1}{2}\,({\bf a}\,{\bf b}\,+\,{\bf b}\,{\bf a})
\,=\,\cos\theta_{{\bf a}{\bf b}}\,=\,{\bf b}\cdot{\bf a}\,, \label{quat}
\end{equation}
where ${\theta_{{\bf a}{\bf b}}}$ is the angle between ${\bf a}$ and ${\bf b}$.
Clearly, this product is a grade-{\it lowering} operation.
It takes two grade-1 numbers, or vectors, and gives back a grade-0 number, or
a scalar.}

Next, let us look at the ``outer product'' between ${\bf a}$ and ${\bf b}$, as defined by Grassmann:
\begin{equation}
{\bf a}\wedge{\bf b}\,=\,\frac{1}{2}\,({\bf a}\,{\bf b}\,-\,{\bf b}\,{\bf a})
\,=\,I\cdot({\bf a}\times{\bf b})\,=\,-\,{\bf b}\wedge{\bf a}\,,\label{nor}
\end{equation}
where ${I}$ (in the modern parlance) is a trivector.
Unlike the previous product this product is
a grade-{\it raising} operation.${\,}$It takes two grade-1\break numbers, or vectors, and
gives back a new directed number of grade 2; {\it i.e.}, a bivector.
Although an abstract entity, the bivector ${{\bf a}\wedge{\bf b}}$ may be visualized as an
oriented plane segment, hovering orthogonal to the vector ${{\bf a}\times{\bf b}}$.

Using the products (\ref{quat}) and (\ref{nor}), the geometric product\break between the two directed numbers
${\bf a}$ and ${\bf b}$ can now be expressed${\;}$as
\begin{equation}
{\bf a}\,{\bf b}\,=\,{\bf a}\cdot{\bf b}\,+\,{\bf a}\wedge{\bf b}\,.\label{gp}
\end{equation}
This product also takes two grade-1 numbers, or vectors, but gives back an entity that is neither a scalar nor a bivector
in general, but rather a quaternion (or a spinor), made out of the grade-lowering operation ${{\bf a}\cdot{\bf b}}$ and the
grade-raising operation ${{\bf a}\wedge{\bf b}}$. To appreciate this, let us express the two components of the geometric
product (\ref{gp}) more explicitly as
\begin{align}
{\bf a}\,{\bf b}\,&=\,{\bf a}\cdot{\bf b}\,+\,I\cdot({\bf a}\times{\bf b}) \notag \\
&=\,\cos\theta_{{{\bf a}}{{\bf b}}}\,+\,\left(\,I\cdot{{\bf c}}\,\right)\,
\sin\theta_{{{\bf a}}{{\bf b}}} \notag \\
&=\exp\{\left(\,I\cdot{{\bf c}}\,\right)\,\theta_{{{\bf a}}{{\bf b}}\,}\}\,, \label{quatty}
\end{align}
where ${{{\bf c}}={{\bf a}}\times{{\bf b}}/|{{\bf a}}\times{{\bf b}}|}$. It is now clear that the product ${{\bf a}\,{\bf b}}$
is a quaternion (or a spinor) that represents a rotation by an angle ${2\theta_{{{\bf a}}{{\bf b}}}}$ about the ${{\bf c}}$-axis.
The geometric product, ${{\bf a}\,{\bf b}}$, thus takes two grade-1 numbers, or vectors ${\bf a}$ and ${\bf b}$,
and gives back {\it a pure act of rotation}.

\smallskip
%\vspace{0.5cm}

\end{document}